\documentclass{llncs}
\usepackage{theorem}
\usepackage{graphicx}
\usepackage{ifthen}
\newcommand{\bracs}{[]}
\newcommand{\CSPM}{\hbox{CSP$_M$}}

\usepackage{epsf}

\usepackage{amsfonts}
\usepackage{epsf}

\begin{document}

\title{Card games as pointer structures: case studies in mobile CSP modelling}
\author{A.W. Roscoe}
\institute{Oxford University Department of Computer Science}
\maketitle
\begin{abstract}
The author has long enjoyed using the CSP refinement checker FDR to solve puzzles, as witnessed by examples in~\cite{tpc,ucs}.  Recent experiments have shown that a number of games of patience (card games for one) are now well within bounds.  We discuss the modelling approaches used to tackle these and avoid symmetric
states.  For two such games we reveal much higher percentages of winnable games than was previously believed.  The techniques developed for some of these card
games~-- which employ various dynamic patterns of cards~-- suggest techniques for
modelling pointer structures in CSP and FDR analogous to those used with the
$\pi$-calculus. Most of these use CSP's ability to express mobile systems.
\end{abstract}

\section{Introduction}
The author has long enjoyed solving mathematical and combinatorial puzzles, so
the advent of FDR in the early 1990's quickly led to him to try some of them
out.  The interested reader will find solutions to peg solitaire in~\cite{tpc,ucs} and to Sudoku in~\cite{ucs}.  The former has become perhaps
the best known benchmark for different versions and implementations of FDR,
for example in~\cite{fdr3,fdr3cluster}. These examples proved very useful for
demonstrating how much better model checking is than humans at finding counter-examples in many situations,
and for teaching students how to use FDR effectively. At the same time
they have been a useful source of student exercises.\footnote{The author gives an annual advanced course in the use of FDR and CSP, and often uses a
puzzle as the first half of the take-home assignment.  Previous years have
included sudoku, magic squares, the unique non-trivial magic hexagon, reversible
peg solitaire, knights exchange (also used as a benchmark in~\cite{fdr3,fdr3cluster}), and ``jumping frogs'' in multiple dimensions.  The 2015 and 2016 ones were based respectively on
on Freecell and Montana solitaire, two of the examples from the present paper.}

At the same FDR has been used in a wide range of practical contexts, and the
methods developed for efficiently coding these and puzzles frequently cross-fertilise.

The author had never, until recently, coded a game of cards in CSP.  This
was for a variety of reasons:
\begin{itemize}
\item The 52! different orderings of a pack of cards seems prohibitive in contemplating potential possibilities.
\item 
CSP's models and the way FDR works do not provide an efficient means for deciding the winner in games with multiple players trying to beat each other.\footnote{The received wisdom has been that they cannot do this at all, however
in writing this paper I have discovered that there is
a contrived way (that would not scale well to difficult problems) of doing it, illustrated the accompanying file
{\tt tictactoe.csp}, which analyses the winnability of the simple game
{\em noughts and crosses} or {\em tic tac toe}.  It shows how the algebra of
CSP can calculate a general {\em minimax} over a tree.}
\item Most games of patience (i.e. card games for one, getting round the previous objection) have dynamic structures, with the cards being
moved around into structures that change size or shape as the game progresses.
FDR has generally been considered as much more suitable for static structures.
\end{itemize}

The first (state-space) objection can be addressed by observing that solving
a {\em particular} deal of patience need not have anything like 52! states,
though plainly the state space sizes of games with dynamic structure are
potentially increased by this in a way not directly related to the number of
deals.  Also, the ever improving capabilities of FDR~\cite{fdr3,fdr3cluster}
(for example parallelising efficiently on multi-core and cluster architectures: the 187M states
of peg solitaire have been brought down to 23 seconds on our 32-core server and less than
2 seconds on a 1024 core cluster)
mean that the sizes of problems that can be addressed routinely using
FDR has increased greatly in the last few years.  In fact virtually none of the
patience game checks that the author has run from the examples quoted here
exceed $10^9$ states, a pretty routine level with the multi-core implementation
of FDR at the time of writing, though the time taken for runs is a little more
than implied by the solitaire figures above, because of the greater complexity
of the CSP models.  Few take more than 5 minutes exploring the state space
 on the
server discussed here.

The second objection is addressed by considering patience games.  The third
is alleviated by the observation that CSP has been shown, for example in~\cite{rpi} and Chapter 20 of~\cite{ucs} to be capable of handling mobile structures.

In this paper we consider the issues of coding patience games in CSP, which
includes building various dynamic structures and avoiding various ways in
which symmetry can multiply state spaces.  Some of these structures are
in effect linked lists and other pointer-based constructions.

This work gives us insight into how to support the translation into CSP
of more general programming models
requiring similar structures, particularly ones based on pointers.

Most readers will be familiar with the card games called {\em patience} or {\em solitaire}, having played them
via programs, apps, websites, or if they are (like the author) sufficiently
old, with real cards.\footnote{The latter (see figures) also proved useful in
checking CSP descriptions.}  We use the name patience in this paper, to provide
a clear distinction from peg solitaire.

A game starts by shuffling one, or sometimes two packs of cards, then dealing
some or all of them into a pre-set pattern.  The player then manipulates
the cards according to the rules of the particular game, usually aiming to
arrange all the cards into ordered suits in some way. A given deal is
winnable if this is a possible outcome. In some cases the question of
winnability is complicated by hidden cards or further nondeterminism
beyond the initial shuffle.  We will discuss this later.

In the next section we give background information about CSP, FDR and a
few games of patience.  Section~\ref{coding} discusses how to code them, together
with our results.

All the CSP programs discussed in this paper can be downloaded from the author's publications web page, along with this paper.

\section{Background}
\subsection{CSP}
In this paper we will make extensive use of the extension of CSP by functional
programming constructs, known as \CSPM, and described in detail in~\cite{tpc}.  This allows to create data types, use inbuilt structures such as tuples,
lists and sets, and to write concise programs which lay out large networks of
component processes indexed by such structures. Thus we can model the various
components of a card game at the programming level, and then, for example,
lay out the initial configuration of a game as a network of one process per card. For this reason the notation given below is that of \CSPM\ rather than the
blackboard language seen in most books and papers. The main CSP operators
used in this paper are:
\begin{itemize}
\item {\tt STOP}\quad The process that does nothing.
\item {\tt a -> P}\quad The process that communicates the event {\tt a} and then
behaves like {\tt P}.   The single event {\tt a} can be replaced by various
choice constructs such as \verb+c!e?x:D+.  This particular one assumes that
{\tt c} is a channel with two data components.  This construct fixes the
first to be the value of the expression {\tt e} and allows the second to
vary over {\tt D}, assumed to be a subset of the type of the second component.
If {\tt D} is absent it varies over the whole of this type.  In cases with such
an input, the successor program {\tt P} can include the input identifier {\tt x}.
\item {\tt P \bracs\ Q}\quad is the external choice operator, giving the choice of the
initial visible events of {\tt P} and {\tt Q}, and then continuing to behave like the one chosen.  Like a number of other CSP operators, \CSPM\ has both the binary and an
indexed form \verb+[] x:A @ P(x)+, meaning the external choice over all the
processes \verb+P(x)+ as \verb+x+ varies over the set \verb+D+.  The effect of
the blackboard CSP $?x:D\rightarrow P(x)$ (which has no literal translation in
\CSPM) can be achieved via \verb+[] x:D @ x -> P+.

$\CSPM$ also has \verb+P |~| Q+ as its internal, or nondeterministic choice operator, which allows the process to choose between the arguments.  However this
operator will not feature in this paper.

\item The final constructs we will use for building sequential processes allow
us to put two processes in sequence: \verb+SKIP+ is a process that terminates
successfully and \verb+P;Q+ lets \verb+P+ run until it terminates successfully
and then runs \verb+Q+.  Successful termination is represented by the process
communicating the special event $\surd$, and is distinct from \verb+STOP+.  Thus \verb+STOP;P = STOP+ and \verb+SKIP;P = P+.

\item CSP has a number of parallel operators.  The ones we shall use here are
{\em alphabetised} parallel and {\em generalised parallel}.  Alphabetised
parallel is written \verb+P [A||B] Q+, where \verb+P+ and \verb+Q+
are respectively communicating in their alphabets \verb+A+ and \verb+B+, with
them forced to synchronise on actions in the intersection of their alphabets.
Most of our uses of this operator will be in its indexed form \verb+|| i:I@[A(i)]P(i)+ where \verb+A(i)+ is the alphabet of \verb+P(i)+, and an action
in the union of these alphabets occurs just when all the \verb+P(i)+ with it
in their alphabet perform it.  Generalised parallel \verb+P [|A|] Q+ specifies
just the interface \verb+A+ that \verb+P+ and \verb+Q+ must synchronise on,
with them being free to communicate anything else independently.

\item Two important operators are commonly used to structure parallel networks:
{\em hiding} \verb+P\X+ allows internal or irrelevant communications \verb+X+
to be hidden, becoming invisible actions $\tau$.   The other is {\em renaming}
\verb+P[[R]]+ where \verb+R+ specifies a collection of pairs of events
written \verb+a <- b+.  For each such pair, every time \verb+P+ offers \verb+a+,
\verb+P[[R]]+ offers \verb+b+ leading to the corresponding state of \verb+P+,
still renamed.  The relational renaming implicitly maps any event of \verb+P+
not otherwise renamed to itself.  It is permissible to map one event of
\verb+P+ to several different ones (which become alternatives)  or several
events of \verb+P+ to the same target.

\item Some of our uses of hiding and renaming will have the effect of creating
a model which can easily be related to a natural specification.  Another
operator we will use for this purpose is {\em throw}: \verb+P [|A|> Q+, in
which \verb+P+ progresses except that if it ever communicates an event in
\verb+A+ it passes control to \verb+Q+.  Thus, like sequential composition,
it represents a way in which \verb+P+ can hand control over to \verb+Q+.
\end{itemize}

\subsection{Some games of patience}
There are a number of books on patience, for example~\cite{solitairebook}, and many web-sites
offering implementations of it.  A number of standard terms
are used. Quoting in part from from Wikipedia\footnote{\tt https://en.wikipedia.org/wiki/Glossary\_of\_patience\_terms}, the following terms
describe collections of cards which exist in some games:
\begin{itemize}
\item{\em Stock}\quad Typically squared and face-down. These cards can be turned over into the waste, usually one-by-one, but sometimes in groups of two or three (depending on individual game rules), whenever the player wishes.
\item{\em Waste} or {\em Discard} or {\em Heap}\quad The area where the cards from the stock go when they are brought into play. The following are typically true:
    The pile is squared and face-up.
    In most games, only cards from the stock can be played to the waste.
    Only the topmost card is available for play.
\item{\em Foundation}\quad This usually consists of the cards build up in individual
suits, face up and in order.  The objective of most games is to move all cards
here.
\item{\em Tableau}\quad The tableau consists of a number of tableau piles of cards. Cards can be moved from one pile or area to another, under varying rules. Some allow stacks of cards that match the building requirements to be moved, others only allow the top card to be moved, yet others allow any stack to be moved.
\item{\em Reserve}\quad A group or pile(s) of cards where building is usually not permitted. These cards are dealt out at the beginning, and used, commonly one card at a time, during the play.
\item{\em Cell}\quad Common to ``FreeCell'' type games, cells allow only one card to be placed in them. Any card can be put in a cell. These act as manoeuvring space.
\end{itemize}

When one is writing a program to test to see if a given deal of a game is
soluble, there are two important issues:
\begin{itemize}
\item In some games all the cards are laid face up on the table, whereas in others some are face up and some are face down or are hidden in the stock.  In the former sort the player
can take everything into consideration before deciding which move to play next,
while in the latter there are things which would have affected the play, but
the player does not have full information without cheating and taking a look.
\item In most games the cards are shuffled before the start but beyond that what
happens to the cards is completely in the hands of the player.  However in some there is re-shuffling or some other element of randomness introduced after the
start.  A good example of this is the game Montana, in which the 52 cards
are dealt out in four rows of 13 and the aces removed.  The game is based on
the spaces this removal creates.  See Figure~\ref{montanapic}.  
\begin{figure}
\begin{center}
\includegraphics[width=20pc]{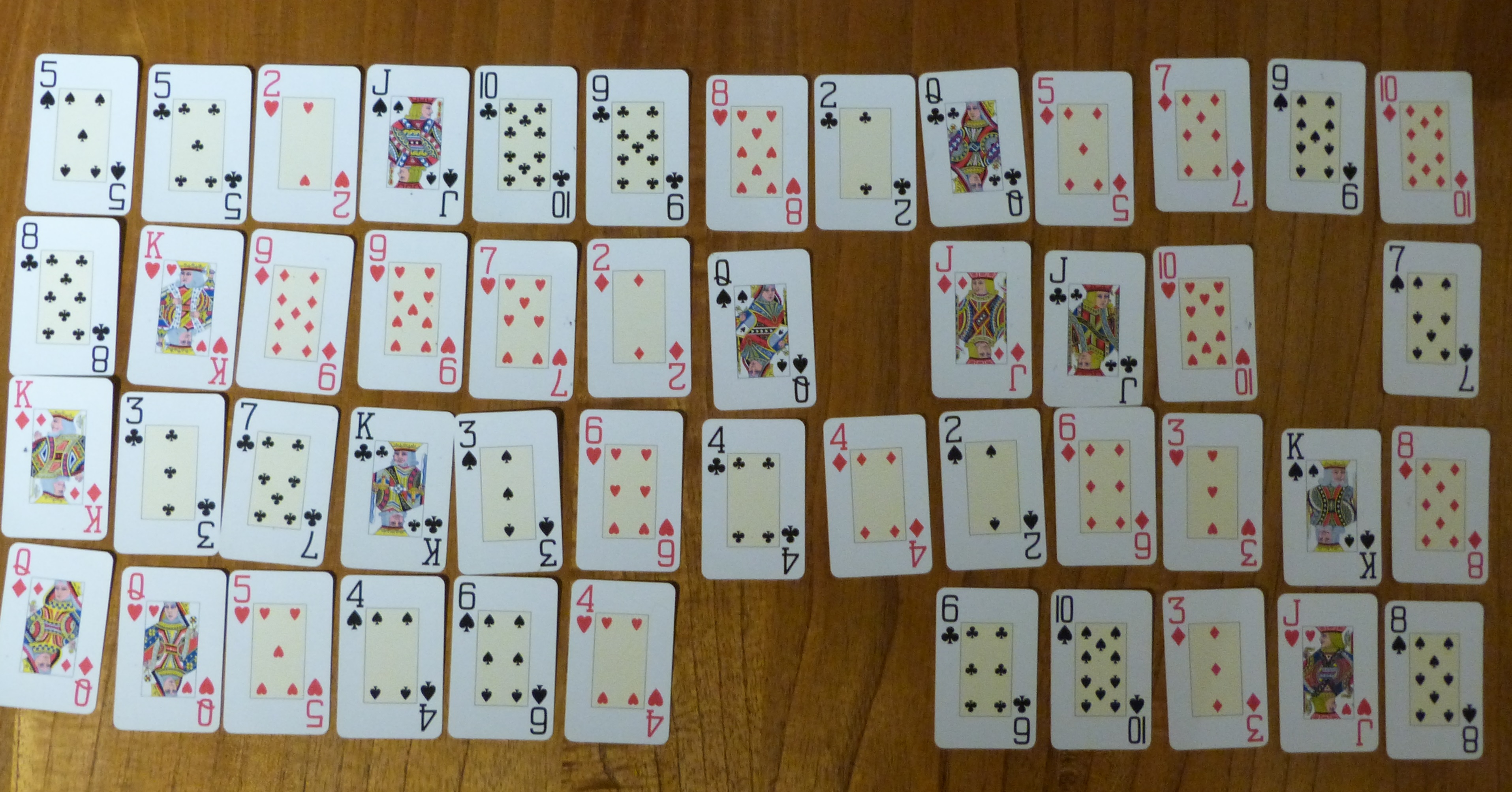}
\end{center}
\caption{A Montana deal after the aces are removed\label{montanapic}}
\end{figure}

A move in Montana is either to place a 2 in a space that appears in the first
column or, in any other column, to place the card which follows (in the same
suit) the card immediately to the left of the space.  Thus the space to the
right of a king cannot be filled and the game becomes stuck when all four spaces
lie to the right of a king with no other card in between.  
In the position shown in Figure~\ref{montanapic} there are three moves possible, namely $K\clubsuit$ moved to after $Q\clubsuit$, $J\heartsuit$ after $10\heartsuit$ and $5\heartsuit$ after $4\heartsuit$.

The objective is
to get the 2-to-king of the four suits in (respectively) the four rows,
in order, leaving a single gap at the end of each row.  Since this is difficult, the game provides that when stuck you can pick up all the cards which are
not already in place, plus the aces, shuffle them, and lay them out in the vacant positions.  The aces are once more removed and the game re-starts.  Usually
three of these re-deals are allowed.

In real life it is unknowable how these re-deals will behave, and so impossible
to condition one's strategy before one of them on how they will turn out.  Note that in this case a re-deal might put all the cards into exactly the right spots
or place each of the aces to the immediate right of a king (so solving or
immediately blocking the game).

We will not consider games of this type, and so in particular will only
consider Montana in the case where there are no re-deals.
\end{itemize}
In the case of games with hidden cards, in seeking a solution to a given deal
we will be finding out if the player could have performed a series of moves
to win the game, which in some cases, where cards are hidden, might be lucky guesses.  

So searching for a solution on a tool like FDR only models what an optimal
player ought definitely to be able to achieve where there are no hidden cards and no
randomisation after the initial shuffle. Aside from no-redeal Montana, two
other such games are Freecell and King Albert, as described here:
\begin{itemize}
\item Freecell is a very popular and well understood game thanks to Microsoft
including it in Windows complete with a method of generating numbered
pseudo-random deals from a seed\footnote{
\tt https://en.wikipedia.org/wiki/Microsoft\_FreeCell}.  The cards are initially laid out in eight columns 
with seven or six cards as shown in Figure~\ref{freecellpic}.  
\begin{figure}
\begin{center}
\includegraphics[width=18pc]{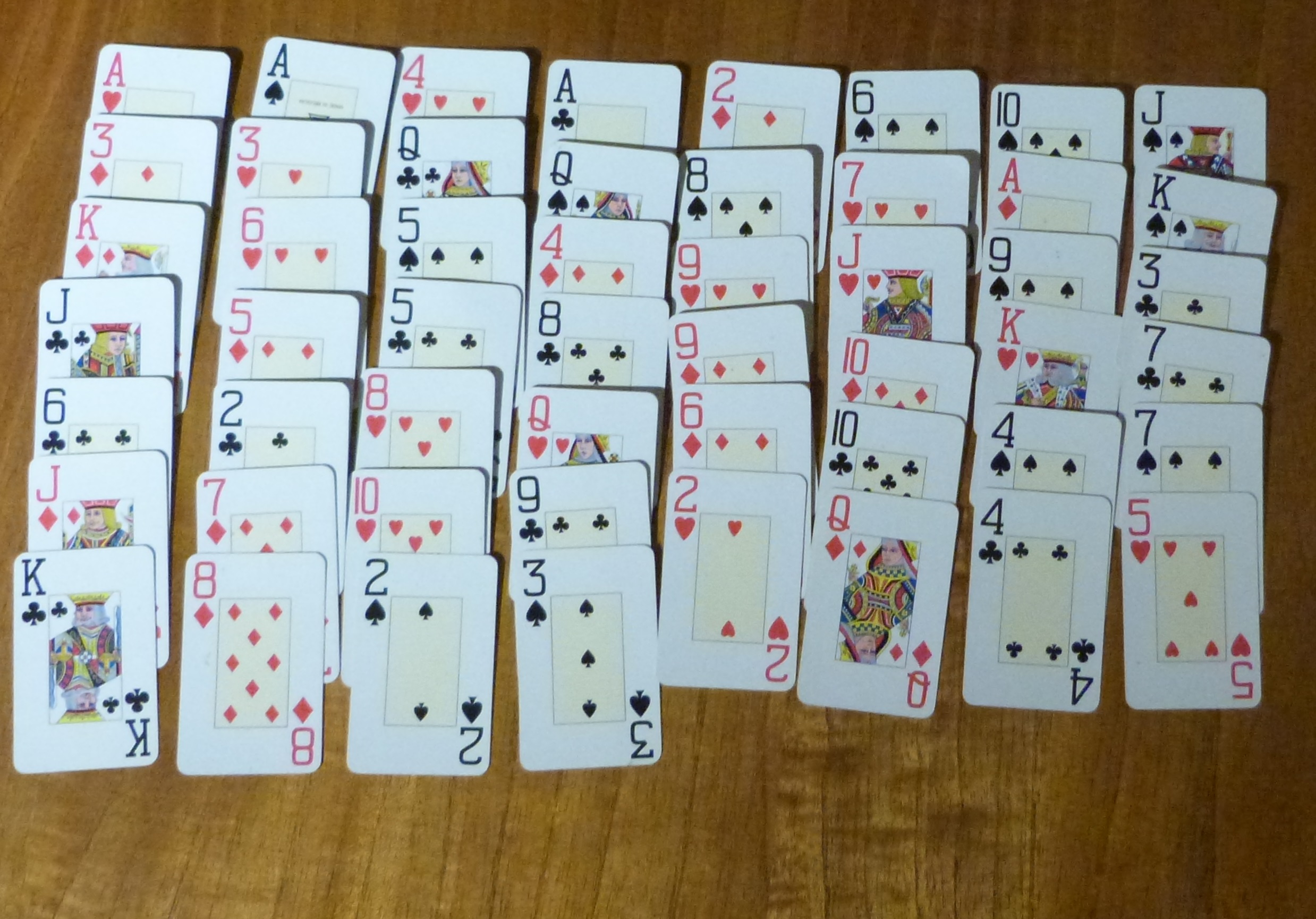}
\end{center}

\caption{Freecell (cells and foundation empty and not shown): not quite the impossible deal 11982\label{freecellpic} (see text)}
\end{figure}

The objective
is to build them up in the foundation as four suits, starting with the aces.
In the basic rules cards are moved around singly\footnote{Most computer implementation allow the moving of stacks of cards with descending alternate colour
from one stack to another {\em but only provided such a move is possible given
the number of cells and empty columns available}.   One could produce versions of the
CSP coding that allowed such extended moves.  This could be done (as seems to be done in
most implementations) by involving heuristics about how many cells and empty columns there are) or potentially by invoking CSP coding to carry out the moves coupled with hiding and
the FDR {\tt chase} operator that eliminates the resulting $\tau$ actions.} between the columns and the
four cells, each of which can hold one card.  A card can either be moved
into the foundations (preserving suit order), into a cell or onto the top
of a column if the top card has one value higher and the opposite colour.
There are no restrictions on what can be moved into an empty column.
Only the top card of each column can be moved.

Thanks to its popularity and the de facto standardisation of deal numbering, Freecell has been well studied with solver programs\footnote{For example {\tt http://fc-solve.shlomifish.org/}}, and a good deal is known about it  including that only 8 of the first million deals are not solvable. The deal shown in Figure~\ref{freecellpic} was the result of the author trying to photograph the first of these, deal 11982.  However there is a mistake: the deal 11982 has $10\heartsuit$ and $8\heartsuit$ swapped from their true positions in that deal.  It is interesting that this small change actually makes the deal solvable (needing the full 4 cells), just as deal 11982 is soluble with 5.  The distributed file {\tt freecelldeal.csp} illustrates these facts.

\item King Albert (named after a king of Belgium) is less well known, and indeed
was unknown to the author until he was working on this paper.  In this the cards are dealt out
into nine columns as shown in Figure~\ref{kapic}, 
\begin{figure}
\begin{center}
\includegraphics[width=18pc]{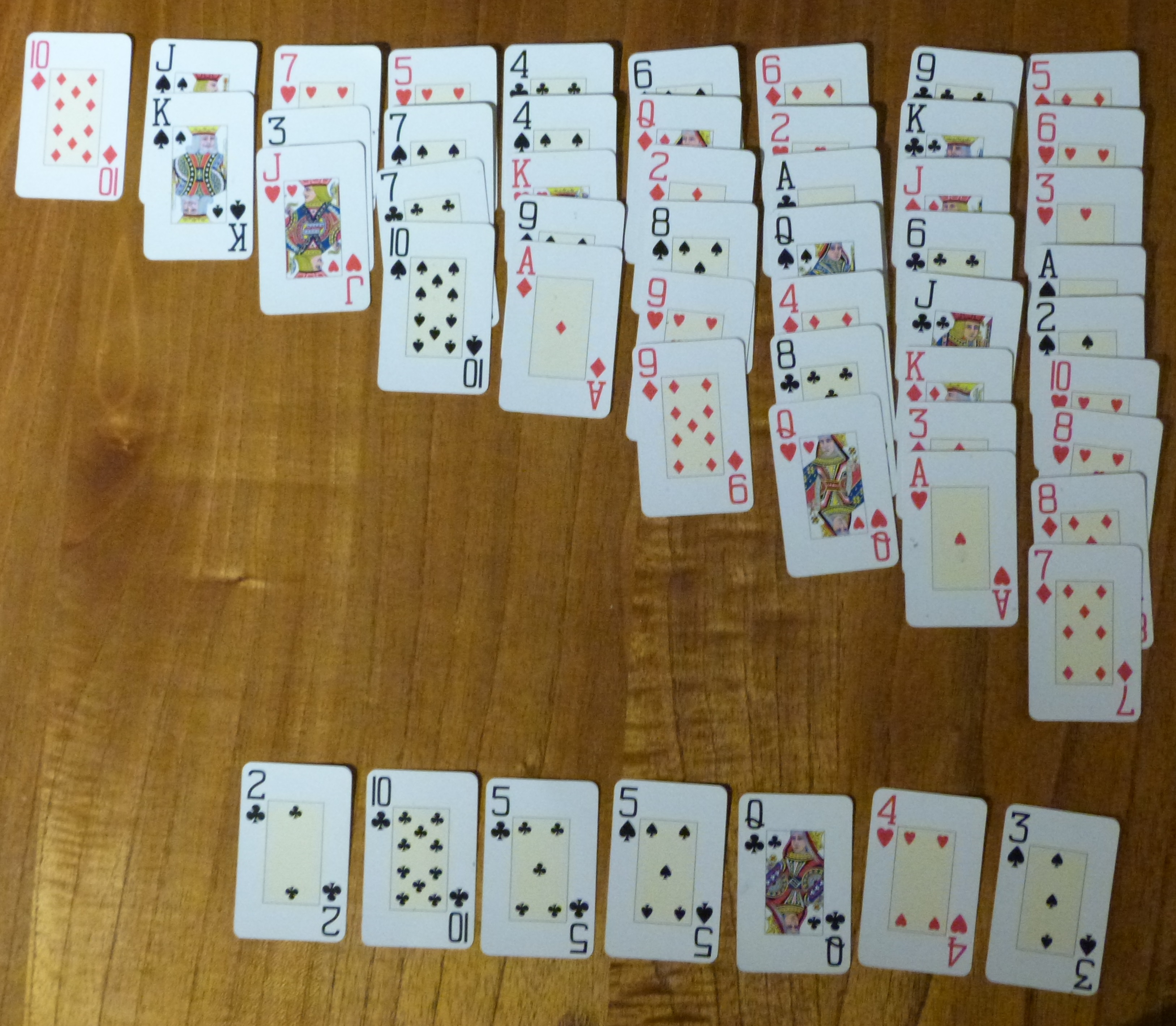}
\end{center}
\caption{A deal of King Albert, showing reserve cards below and foundation missing and empty\label{kapic}}
\end{figure}
leaving seven cards as the
``Belgian Reserve''.  This game is played exactly like Freecell except that
there are no cells, and cards can only be moved from the reserve, never into it.   The author has found no evidence of prior computer analysis of this
game, merely a frequently repeated statement that about 10\% of games are solvable (we will see later that this is wrong).
\end{itemize}
Nor has the author found any evidence for computer analysis of Montana, though
one does often find the statement that about 1 in 20 games or 1 in 10 games is soluble when
using the three redeals.  We will see later that this is also wrong.

\begin{figure}
\begin{center}
\includegraphics[width=18pc]{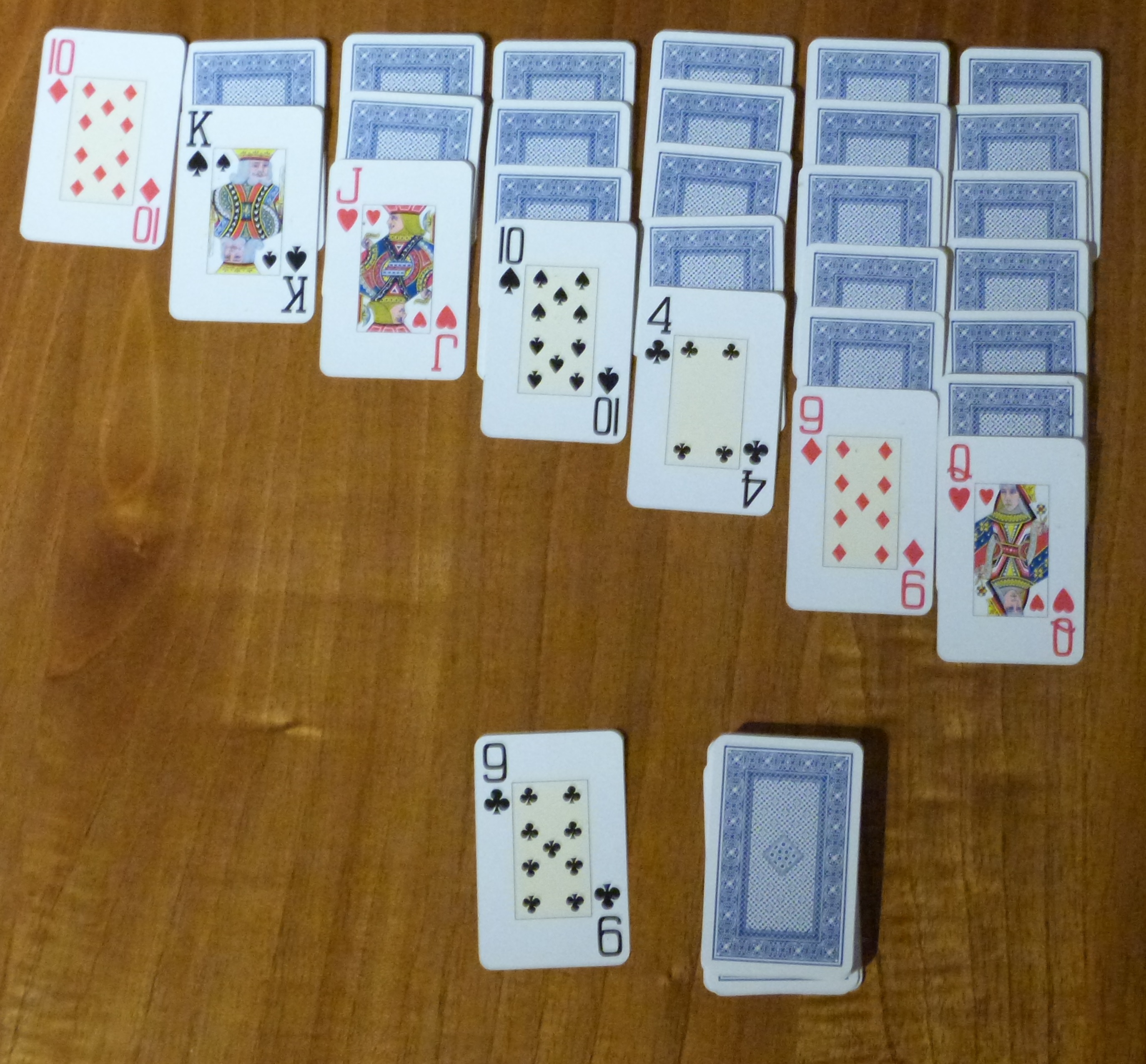}
\end{center}
\caption{Deal of Klondike, with stock and waste below, foundation missing and empty\label{klondikepic}}
\end{figure}

The final game we consider is perhaps the best-known one, Klondike~\footnote{http://en.wikipedia.org/wiki/Klondike\_(solitaire)}: see Figure~\ref{klondikepic}
  Unlike the ones above this
does have hidden cards (initially only eight are visible, the tops of the
seven columns and the top card of the 24-card stock.  It is played like
King Albert except that a card in one of the columns is only turned face-up
when it is at the top, stacks of alternating descending cards can be moved
freely between columns provided the alternating descending property in maintained, but only a stack with a king at the bottom can be moved into an empty stack.
Usually (unlike Freecell and most descriptions of King Albert) cards can be
moved from the foundations to the columns, preserving the alternating
descending property.

There are many variants on this game, mainly based on how the stock cards
are dealt (typically singly or in threes, with various numbers of times through
the stock).  There are figures for the solubility of this game (based on perfect knowledge)
of several variants\footnote{\tt https://en.wikipedia.org/wiki/Klondike\_(solitaire) }.  In this paper we will concentrate on the case where cards are dealt singly, with no limit on the number of times a card can
be turned over.

\section{Modelling patience in CSP}\label{coding}
In CSP, particularly when this is designed for input into FDR, we want models
to have a fairly high degree of parallelism, since the tool is typically more
efficient for these.  We also want to avoid creating
any inessential states, which in most games means states that do not relate
directly to positions where real moves have been completed.  For if we model a
single move by several, especially if we allow these to interleave, the state space
can grow alarmingly.  

All of the games we have described have symmetries which mean there can, depending on
how they are coded, be large
classes of states which are completely equivalent in behaviour.  In the three
examples based on columns of cards, it is clear that two states with the same
set of columns (though in different positions) are completely equivalent
as regards solubility, as are two positions obtainable from each other by
permuting the rows in Montana.  

Aside from Montana, which has an essentially fixed format throughout (four rows
of 13 cards or spaces), all the others change in pattern dynamically through
the game as various piles and stacks of cards grow, shrink and are moved around.

There are two obvious approaches to modelling these games in CSP.  One is to
make each card a process, which at any time knows enough about its position
to allow it to determine its ability to move or to be moved to.  These cards,
together with one or two other processes to monitor things like empty cells or
columns, can make up the CSP model of a game.  The other is to model each component of the
game such as a column, an individual suit in the foundation, the stock or
the waste cards.  In the case of Montana it makes sense to model each position in the $4\times 13$ grid as a separate process that is either a card or is
a space (this state changing as moves are played).  Each of these has its
advantages and disadvantages, as revealed in the following discussion.

In the card-based solution, each card needs to know when it is at the top of
a column or otherwise available for moving.  It therefore has to know not
only when it is moved, but also whether a card moves from or to being on top of it.  This means that each card needs to know what card is above it and/or which is
below it, noting that when a card is at the top of a column there will be none
above, and none below when on the bottom.

The author dealt with this by using the first of the following data types:
\begin{verbatim}
datatype links = Crd.card | Top | Bottom

datatype suits = Hearts | Clubs | Diamonds | Spades

datatype card = Card.suits.{1..13}
\end{verbatim}
making each column into a linked list of processes.   There are arguments for making this
doubly linked (each card knowing the identities of the cards above and below it), though
in many circumstances one can  get away with a singly linked list. However in the
latter case it is  still for a card to know if it is at the top
of a column, and probably necessary for it to know if it is at the bottom of one, so that moving it creates an empty column.

He used actions such as \verb+stacktostack.c1.c2+ meaning that
\verb+c1+ moves from the top of a column to the top of another\footnote{Where there is only one card that {\tt c1} can move on top of, as when it is moved
to the top of a suit, there is no need for the parameter {\tt c2}, and where there is a restricted range of targets for {\tt c1} we can again reduce the
range of {\tt c2}.  In the common case where cards are only moved on top of
one of the opposite colour and one point higher, we can replace {\tt c2}
by a suit.}, where the current top card is \verb+c2+.   Clearly the processes representing \verb+c1+ and 
\verb+c2+ respectively need to agree to this move, with \verb+c2+ agreeing only
if it is at the top of a column and \verb+c1+ agreeing only when it is at
the top or in the sorted portion in the case of Klondike.  Such moves will only
be possible when \verb+c1+ is of the opposite colour and one point lower than
\verb+c2+.

The the cards now at the top of the two stacks will have to know this.  It follows that whatever card was below \verb+c1+ will need to synchronise on the event
also or otherwise be told about its new status.

Because a card keeps the same alphabet at all times, but the card on top of
it changes, its alphabet (in the case where it synchronises) will have to contain all moves of the form
\verb+stacktostack.c1.c2+ where \verb+c1+ is any card which might ever be
on top of it.  Furthermore it will have to agree to such events: in the case
where the card \verb+c+ we are defining is \verb+c2+, it will only agree to this event
when it is on top of a stack.  If it is not \verb+c2+ but \verb+c1+ is on
top of it, \verb+c+ knows that after the event it is on top.  However (wherever
it is in the game) \verb+c+ should never block the event \verb+stacktostack.c1.c2+ when it is neither \verb+c1+ nor \verb+c2+.

This is similar to the techniques used in Chapter 20 of~\cite{ucs} to model mobility,
where processes move around and change which ones they communicate with.  There
the author showed that this could be simulated in CSP by putting every event that ever be available to a process into its simulation alphabet, and having 
the simulation of each process that
does not for the time being have an event $a$ in its mobile alphabet agree
to that event.  The key way of explaining this is the CSP identity
\begin{verbatim}
P [A||B] Q  =  (P|||RUN(B'))[|Events|](Q|||RUN(A'))
\end{verbatim}
where \verb+A'+ are the members of \verb+A+ not in \verb+B+ and vice-versa,
provided \verb+P+ and \verb+Q+ do not terminate with $\surd$ and only communicate within \verb+A+ and \verb+B+ respectively.  Thus the parallel composition
on the left hand side, with its fixed alphabets, is equivalent to the one on
the right where a pair of processes always synchronise on everything after
being padded out by the two \verb+RUN(X)+ processes, which are always
willing to communicate anything in the parameter set of events.  By building
something like a dynamic version of \verb+RUN(X)+ which changes as its
accompanying process's alphabet changes, we can create the effect
of \verb+P [A||B] Q+ where \verb+A+ and \verb+B+ can change as the 
execution progresses.  That is precisely the approach set out in   
\cite{ucs}, whereas here we take the approach of building the
process representing each card so that it allows the extra events required
for mobility itself by allowing but ignoring events that would be
relevant only if the moving card is in a different position.

So the state of a card when it is in one of the eight stacks of Freecell, the
first game we consider here, is
\begin{verbatim}
CStack(c,below,above) = above==Top & stacktosuit.c?_ ->
  (NewTop(c,below); CSuit(c))
[] above==Top & stacktostack?c':skabove(c)!c -> CStack(c,below,Crd.c')
[] above==Top & stacktofc!c -> (NewTop(c,below);CFC(c))
[] above==Top & fctostack?c':skabove(c)!c -> CStack(c,below,Crd.c')
[] above==Top & stacktostack!c?c':skbelow(c) ->
                    (NewTop(c,below);CStack(c,Crd.c',above))
[] IsCard(below) and above==Top & stacktoestack.c ->
                   (NewTop(c,below);CStack(c,Bottom,above))
[] above!=Top & maketop.cardof(above)!c -> CStack(c,below,Top)

NewTop(c,c') = if c'!=Bottom then maketop.c.cardof(c') -> SKIP 
                             else freestack.c -> SKIP

LSbelow(Card.S.n) = if n==Ace then Bottom else Crd.Card.S.n-1
\end{verbatim}
Note that there are separate events for the possible
places for moves to and from a stack such as a cell or suit.
The separate event for
moving to an {\em empty} stack is because this is governed by different
rules to moves to a non-empty stack.  There is a separate processes
which monitor the numbers of empty stacks 
 and only allows a card to
move into one  when one is available.

It is important to realise that a card has no idea which of the eight stacks
it is in.  That is because it really does not matter: all that matters is
that it is in a stack and which cards are above and below it.  Note that if
we did have an index for which stack a card is in, this might increase the
state space by a factor of nearly $8!$,  because any permutation of the
stacks would now produce a different state except for swapping two or more
empty ones.\footnote{In practice one cannot swap stacks that are not completely
sorted into decreasing alternating order, but in soluble games there are
many possible permutations of all-sorted ones.  Not multiply counting these
makes a huge difference to the state space of soluble games.}  This is
perhaps the strongest reason why we code stacks as clusters of mobile
card processes rather than a single process that represents a given stack,
although another strong one is that such processes would have large
state spaces.\footnote{Given that aces can be assumed never to
be in a sorted stack, there are exactly $32,712 = 2^{12+3}-12*4 -8$ sorted ones.This does not take account of the unsorted cards that may be in the stack
above these.}

Similarly there is a very simple state for a card in a cell, with
cards entering these when allowed by a separate process that monitors how
many are currently filled.  Just as with stacks, there is good reason not
to let each card know {\em which} cell it is in.
\begin{verbatim}
CFC(c) =
fctosuit.c?_ -> CSuit(c)
[] fctostack.c?c':skbelow(c) -> CStack(c,Crd.c',Top)
[] fctoestack.c ->  CStack(c,Bottom,Top)
\end{verbatim}

There is similarly a process which monitors the progression of cards into
the four suits, ensuring that each suit is built up in order and also raising
a flag when all four are complete.  It has a further, more subtle purpose.
If a card is available to put in a suit then it is not always optimal to put
it there, because it may be needed to hold a lower card of the opposite
colour on top of it in a stack.  However if both the cards of opposite colour with value one lower are in the suits {\em or} both cards of opposite colour
with value two lower {\em and} the card with value three lower and the same
colour (but different suit) as $c$ are already there then there can be no point in not moving $c$
to its suit.\footnote{The first of these cases is obvious: the only two
cards $c'$ that $c$ might hold are already in the suits and cannot move back.  The
second case follows because under these circumstances it is possible and
 safe to move any such $c'$ to the suits: there is no need for $c$ to
hold it.}

The Boolean flag which is part of the event moving each card to its suit is
calculated by the \verb+Suits+ process: it is true is it is safe to force
the move according to the above criteria.  

We force such actions to happen when they can.  By itself this considerably
reduces the state space, but we have another trick to play as well.  We
hide forced events and apply the FDR {\em chase} operator which forces
{\em tau} events to happen without counting them as events in the breadth-first search.  This has the effect of bringing forward states in which a lot of
moves to suits have been forced, meaning that FDR typically looks at a
lot less other states before finding one of these than it would without
{\em chase}.  In fact if the forcing is made optimistic: forcing all enabled
moves to suits, this usually results in finding solutions faster, though
in theory it could fail to find any solution when there is in fact one.

The events \verb+freestack.c+ indicating that a stack has become empty
are also hidden and {\em chased}, because it is safe to do so and also
reduces the state space by eliminating interleaving.

There are several files implemented using these ideas available with this
paper on the author's website.  These include ones with a random deal generator
that requires a seed, one that cannot be solved with four cells,
and one that uses suits rather than cards as the second argument of 
\verb+stacktostack+ as discussed above.  At the time of writing 
all 52-card deals attempted were within the range of the author's dual-core
 laptop if not offered
more cells than they need, most taking a  few minutes\footnote{To illustrate this, for
an arbitrarily chosen random deal the author tried, 0 cells resulted in a model with
only a few states and no solution, 1 cell led to a solution on ply 59 of the FDR search and
227K states, 2 cells found a solution in 40 plies but used 58M states, 3 used 34 plies and 257M states, while 4 used 31 plies and 734M states. The decrease in the number of plies with more cells is because these give the player more move options.}

A feature common to all our card-game models is that FDR's preliminary
compilation phase in which it reduces a CSPM script to the supercombinator
form the checking run in, takes significant time for each deal irrespective of
how many states that deal has.  In the case of these Freecell models, it
takes about 20s per script on the author's laptop, but this would have
been considerably more if we had used processes representing multiple
cards (like whole stacks) or had use a larger alphabet (for example by
parameterising \verb+stacktostack+ with three cards (the card moving and
the from and to ones), which is very tempting. 

\subsection{Other games}
The coding of King Albert is very similar to that of Freecell: the only
differences are that there is now one more stack, that these have different
sizes, and that the seven-card reserve (to which cards cannot move) replaces
the cells.  This presents no challenge.  The main point of interest in
this game is the results obtained.  Despite the statements on the web to the
effect that about 10\% of games are soluble, the author found that 72 of the
first 100 random deals were solved, mostly fairly quickly, by FDR.

Klondike is similar to King Albert.  The fact that some of the cards in
columns are initially face down does not affect the coding, only the
interpretation of what a discovered solution means~-- as discussed earlier
this now depends on some luck as well as careful planning.  The reserve of
King Albert has now become a 24-card  stock that are initially face down
and turned over in groups of 1 or 3 into the waste.  There is a sense that
a stock that is turned over in 1s as often as the player chooses is equivalent,
in terms of solvability, to a reserve: each card remaining in the stock can
always be reached. On the other hand this would move away from the spirit of
the game and not be adaptable to more limited access to the stock.
The author therefore implemented the combination of stock and waste as
a doubly linked circular list: the links in one direction enable us to turn over the cards from the stock in the correct order, and those in the other enable
us to treat the waste as a stack.  A header process was included in the list
to detect the top and bottom of the stock, handle the process of starting a
new pass through the list, and handle the case where the stock as become
empty.

Cards are never added to this circular list, but we have to allow for
a card $c$ being removed, meaning that the two adjacent processes need to have
the pointers which formerly pointed to $c$ being altered. Thus the event(s)
constituting the move of $c$ to a column or suit need to tell the two
adjacent cards what $c$'s former links were.  It would be preferable to
to do this in a single action, but that would mean that the events would
need to contain at least three cards as data fields, meaning that the
alphabet size and consequent compilation time would become problematic.
Therefore the author actually coded these interactions via a series of
events, using a separate process \verb+Linker+ to avoid having three-card
events. The events are
\begin{enumerate}
\item The main move of a waste card \verb+c+ to a column or suit, that card then
\item communicates its previous \verb+below+ link to \verb+Linker+,
\item followed by its previous \verb+above+ link.
\item \verb+Linker+ then tells \verb+above+ and \verb+below+ to link to each
other in place of \verb+c+.
\end{enumerate}

The last three events are hidden and the FDR function \verb+chase+ applied.
This forces the three $\tau$s created
from the \verb+Linker+ events to happen immediately after the main move,
thereby making the four events effectively into one.  This cuts down on
potential interleavings and eliminates any possibility of error arising from
the series of events relating to different game moves overlapping.

It would be easy to adapt the resulting CSP model to any of the usual variations
in how the stock is used.

The solvability of Klondike has already been well studied (see Wikipedia article),
so the author did not seek to repeat these experiments.

Unlike the three games we have discussed above, Montana maintains a rather
static structure throughout the game: four rows made up of 13 cards and
spaces each, with four spaces and 48 cards overall. Therefore we now have
the option of maintaining a process for each of these 52 slots (which is
at all times a named card or an unnamed space) as well as the option of
having a mobile process for each card (and perhaps space too).  While the
mobile cards approach has the potential benefit of eliminating any symmetries
which arise from permuting the rows, provided one does not create more
symmetries by giving the four spaces separate identities, the disadvantages
mean it is not practical.  In summary these are:
\begin{itemize}
\item A process for a slot in the array only needs 49 basic states (one for each
card that might be in the slot and one for a space.  However to get the
linked list approach to work the list needs to be doubly linked (i.e. each
card has a pointer to both left and right).  It also needs to know how many
spaces (0--4) are on its right, making $48\times 48 \times 5$ basic states
for a card process, and each of these states needs to participate in a large number
of events.  This is significantly more expensive than in the other
examples we have seen of doubly linked lists, because for example the only cards that
can be involved in the stock and waste in Klondike are those initially in the stock.
\item To maintain the state of these pointers and counters at the sites
both where the card is moved from and to needs multiple communications and
large alphabets. Multiple communications lead to a large number of
intermediate states. 
\end{itemize}
These mean that the time taken to compile, run and debug the checks are
substantially greater than with the array model.  Despite using tricks such as factoring the card process into three and using
a lot of renaming to avoid symmetric compilations, the author did not succeed
in getting the mobile card model to run for more than a 20 card pack (once
the aces had been removed), which is the smallest possible with four suits where one cannot
get a complete row of spaces.

On the other hand the array model was very successful and demonstrated that
Montana (not allowing any redeals) is far more frequently soluble than appears
to have been believed before.   63 of the first 100 random deals were soluble
with no redeals, whereas several web sites quote the figure of about 10\% being soluble by a skilled player with three redeals.

\section{Conclusions}
We entered on this study to illustrate the expressive power of CSP, in particular in describing mobile systems, and the power of FDR to decide things about
them.  It is clear that these 52-card games of patience are within the
scope of what can be handled by these
technologies, but are sufficiently close to the boundary that we have
to be careful about how to describe each game, taking careful consideration
of the nature of each system from the points of view of alphabet size,
compilation time and running time.

We have also illustrated a style of describing mobile systems in CSP that
is different from that described in Chapter 20 of~\cite{ucs}, where 
it was shown how to achieve the effect of communicating channels along
channels.  The approach used here is more explicit: we simply tell each
process (i.e. card) who it is next to, keeping this information up to date,
and the processes duly communicate with their neighbours.  Thus processes
are told the topology of their local network explicitly rather than
implicitly.  

Many of the models we created included linked structures of CSP processes
strongly suggestive of the pointer structures in heap used in programming.
We will develop this idea by showing in a subsequent paper how such
programming styles can be supported by CSP and FDR, particularly when
new {\em symmetry reduction} functionality of FDR is used to eliminate the
differences in states created by different choices of heap cells for the
same roles.  That issue is essentially the same as the reason it
was so much better to use the card-based linked model for games like
Freecell rather than to park cards in particular named columns or cells.
The same symmetry reduction technology would successfully eliminate
these symmetries in card models too, but the models presented here do
not require it. 

The main reason for this is that each member of the linked structures we
build is tied to an immutable and distinct value, namely its value as a
card, whereas cells in a heap are identical.

One could well imagine producing a compiler from a suitable language for
describing games of patience to CSP, if one were wanted.  However one might well
need to input some human judgement about what style of model (say array or mobile) would work best.

The CSP files relevant to this paper can be found in an accompanying zip file
on the author's web site.  Note that the one for Klondike incorporates a
trick not discussed in this paper: this uses the ability of FDR to
compute multiple counter-examples to determine, in a single run, what the highest score attainable with a given deal it.  The style of programming for this
exactly parallels one discussed for metering timing problems in Chapter 14 of~\cite{ucs}.

\end{document}